# Integrated magnonic half-adder

Qi Wang[1], Roman Verba[2], Thomas Brächer[1], Florin Ciubotaru[3], Christoph Adelmann[3], Sorin D. Cotofana [4], Philipp Pirro[1], and Andrii V. Chumak[1,5*]

[1]*Faculty of Physics and Research Center OPTIMAS, University of Kaiserslautern, Kaiserslautern 67663, Germany*

[2]*Institute of Magnetism, Kyiv 03142, Ukraine*

[3]*Imec, Leuven, 3001 Belgium*

[4]*Department of Quantum and Computer Engineering, Delft University of Technology, Delft, 2600 The Netherlands*

[5]*Faculty of Physics, University of Vienna, Boltzmanngasse 5, A-1090 Vienna, Austria*

Spin waves and their quanta magnons open up a promising branch of high-speed and low-power information processing. Several important milestones were achieved recently in the realization of separate magnonic data processing units including logic gates, a magnon transistor and units for non-Boolean computing. Nevertheless, the realization of an integrated magnonic circuit consisting of at least two logic gates and suitable for further integration is still an unresolved challenge. Here we demonstrate such an integrated circuit numerically on the example of a magnonic half-adder. Its key element is a nonlinear directional coupler serving as a combined XOR and AND logic gate that utilizes the dependence of the spin-wave dispersion on its amplitude. The circuit constitutes of only three planar nano-waveguides and processes all information within the magnon domain. Benchmarking of the proposed device is performed showing the potential for aJ energy consumption.

*142 words from 150*

A spin wave (SW) is a collective excitation of the magnetic order in magnetic materials that can propagate in both conducting and insulating media. Spin waves and their quanta magnons are considered as a potential data carrier for beyond Moore computing [1-4]. This is due to their ultrashort wavelengths in the micro- to nanometer range [5-7] in the microwave and THz frequency range [8,9], due to their ultralow losses and the absence of Joule heating [10,11], resulting in long-propagation distances of spin information [12,13], as well as because of their abundant nonlinear phenomena [14-16]. These factors make spin wave highly attractive for wave-based computing concepts [17-27]. Separate spin-wave logic gates [17-19], wave-based majority gates [20-23], transistors [15,24,25] and building blocks for unconventional computing [26-29] have already been proposed in experiments or micromagnetic simulations. Thus, the next step is the realization of integrated magnonic circuits consisting of at least two elements that are suitable for further integration and interconnection with other magnonic components. In general, there are two approaches to realize magnonic circuits. One is the development of magnonic circuits using magnetoelectric cells [30-32] and modern spintronics structures [4,33-35] that behave as transducers converting information between magnons and electrons. This concept suffers from a large number of required conversions from spin to charge and vice versa, which have been identified as a serious bottleneck, in particular due to the relatively low conversion efficiencies achieved so far [30]. The second approach is based on the development of all-magnon circuits in which one magnonic gate is directly controlled by the magnons from the output of another magnonic gate without any intermediate conversion to electric signal [15].

---

[*] Corresponding author. Email: andrii.chumak@univie.ac.at

Here, we demonstrate such an integrated magnonic circuit on the example of a half-adder built from two logic gates by means of micromagnetic simulations. The magnonic half-adder has a strikingly simple design, consisting of two directional couplers [36]: The first one works in a linear regime and acts as a symmetric power splitter for each of the two inputs. The second coupler operates in a nonlinear regime and simultaneously performs XOR and AND logic operations.

**Magnonic half-adder design.** A general schematic layout of a half-adder in electronics is shown in Fig. 1a. It combines an XOR logic gate and an AND logic gate using three-dimensional bridge constructions. The half-adder is an essential primary component of any arithmetic logic system, which is why we chose it as proof of the concept device for an integrated magnonic circuit. It adds two single binary digital Inputs "*A*" and "*B*" and has two Outputs, sum ("*S*") and carry ("*C*"). The truth table of a half-adder is shown in Fig. 1b.

The design of the proposed magnonic half-adder is not a direct reproduction of a complementary metal oxide semiconductor (CMOS) half-adder in the spin-wave domain. It uses advantages proposed by SW physics and is based on the combination of two directional couplers [36-38] operating in different regimes and performing different functionalities. The sketch of the device is shown in Fig. 1c. Directional Coupler 1 in the magnonic half-adder acts as a power splitter for each of the two inputs and, at the same time, replaces the three-dimensional bridge [36] required for sending the signals from Input "*A*" to the gate AND and from Input "*B*" to the gate XOR (see

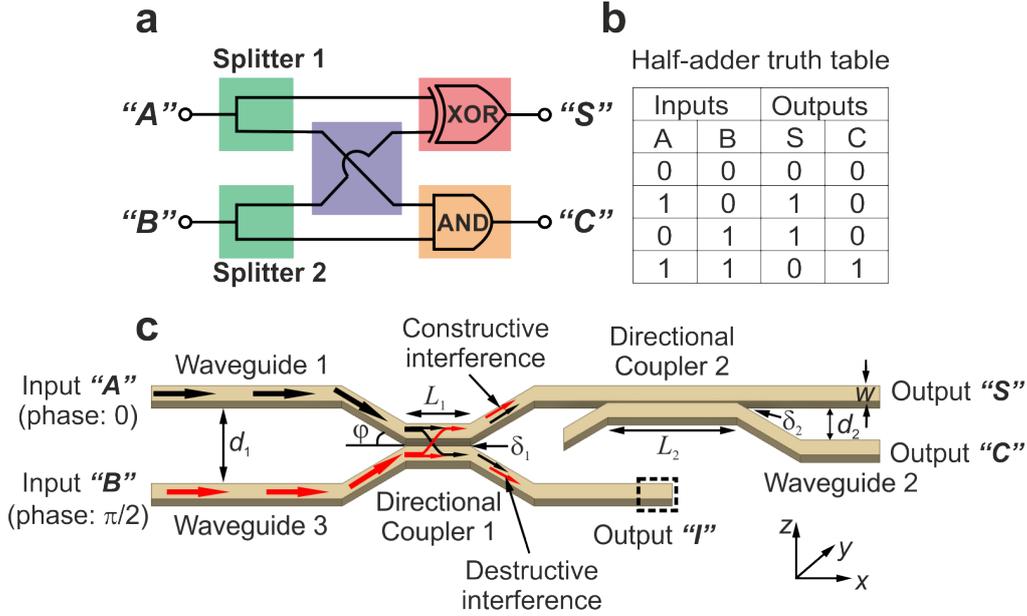

**Fig. 1 The operational principle of the magnonic half-adder**. **a** Sketch of half-adder in electronics. Building blocks are highlighted by different colors. **b** Half-adder truth table. **c** Schematic view of the magnonic half-adder. In this work we consider the following parameters: The widths of the YIG waveguides are $w = 100$ nm, thicknesses are $h = 30$ nm, edge to edge distances between different waveguides are $d_1 = 450$ nm, $d_2 = 210$ nm, the angles between the waveguides are $\varphi = 20°$, the gaps between the coupled waveguides are $\delta_1 = 50$ nm, $\delta_2 = 10$ nm, and the lengths of the coupled waveguides are $L_1 = 370$ nm and $L_2 = 3$ μm. The arrows show the magnons flow path from inputs to logic gates.

Fig. 1a). The spin-wave flow paths in the magnonic half-adder are shown by the black and red arrows in Fig. 1c: Spin waves from both inputs are split into two identical spin waves of half intensity by the Directional Coupler 1. One pair of the waves is directed to the Directional Coupler 2 via the Waveguide 1 and the other pair is guided into the idle Output "*I*" via the Waveguide 3. In the present simulation, the Output "*I*" just features a high damping region at the end (shown in the figure with a dashed rectangle) and it does not contribute to the half-adding function. However, it acts as an XOR logic gate and, with the use of another directional coupler, can perform the same half-adder

operation (see Supplementary Materials). Thus, the modified half-adder can be considered as a combination of a half-adder with a fan-out logic gate which doubles each output of the device. The Directional Coupler 2 performs the actual half-adder operation and its operational principle is described in the next section.

The material we use for the numerical simulations is Yttrium Iron Garnet (YIG) which features the smallest today's known SW losses together with large velocities of short-wavelength exchange spin waves [6,10,11,13]. One of the greatest advantages of magnon-based data processing systems is their scalability, since the smallest wavelength is limited by the lattice constant of the used magnetic material. In our simulations we have chosen the minimal width of the waveguides to be 100 nm (see Fig. 1 for the sizes of the structure) which can be reliably fabricated using modern patterning techniques [39-43]. Recently, the propagation of coherent spin waves in nanoscale waveguides of the width down to 50 nm was studied analytically, numerically and experimentally in Ref. [41,42]. Moreover, the nanoscale directional coupler, which is the primary building block of the magnonic half-adder, has been fabricated and studied using the micro-focused Brillouin light scattering in linear and nonlinear regimes [43].

**Theoretical model of the nonlinear directional coupler**. The processing of data, in general, requires the utilization of elements with nonlinear characteristics that are, e.g., provided by a semiconductor transistor in CMOS. Spin waves possess a variety of natural and very pronounced nonlinear phenomena [44] that potentially can be used for computing such as four-magnon scattering, which was, for instance, used in the demonstration of the first magnon transistor [15]. Here, we utilize the phenomenon of the nonlinear dependence of SW frequency on its amplitude [44,45]. The increase in the SW amplitude results in a shift of the SW dispersion and in a phase accumulation due to the change in SW wavelength while conserving its frequency. The utilization of this phenomenon for data processing has the advantage that, unlike in the magnon scattering-based approaches, no magnon energy is lost to idle magnons generated due to the scattering [15].

The basic configuration of the directional coupler consists of dipolarly-coupled straight parallel waveguides and of bent waveguides in order to guide spin waves in and out – see Fig. 2a. Directional couplers operating in the linear regime were comprehensively studied e.g., in optics and in magnonics [36,37]. A power dependence of the characteristics of SW-based directional couplers was observed in experiments and simulations on mm-scale [16,38] and nanoscale samples [43].

When two parallel magnetic SW waveguides are placed sufficiently close to each other, the dipolar coupling between them results in a splitting of the dispersion curve of the isolated waveguides into symmetric and antisymmetric modes of the coupled waveguides [36,37]. The analytically calculated [36] dispersion relation of the isolated SW waveguide is shown in Fig. 2b by the grey solid line for the case of the lowest SW mode with a quasi-uniform profile across the waveguide width [41]. The split dispersion relations in the linear regime in the coupled waveguides are shown in the figure by the blue lines. To obtain the linear dispersion, small SW amplitudes are excited by a microwave field of $h_{rf}$ = 2 mT. When the excited SW frequencies are above the minimum frequency of the antisymmetric mode (about 2.278 GHz), the symmetric and antisymmetric SW modes with different wavenumber $k_s$ and $k_{as}$ will be excited simultaneously in the coupled waveguides. The constructive and destructive interferences of these two propagating SW modes results in a periodic energy exchange between the two waveguides. In doing so, the spin waves excited in one of the waveguides transfer their energy to the other after propagation over a certain distance which is called the coupling length $L_C$. The coupling length $L_C$ can be calculated as $L_C = \pi / \Delta k_x = \pi / |k_s - k_{as}|$ and depends on various parameters such as the SW wavelength, the applied magnetic field, the geometrical sizes of the waveguides and their magnetizations [16,36-38].

The output power in the first waveguide normalized by the total power $P_{1out}/(P_{1out}+P_{2out})$ can be expressed using the characteristic coupling length $L_{C2}$:

$$\frac{P_{1out}}{P_{1out} + P_{2out}} = \cos^2\left(\pi L_2 / (2 L_{C2})\right), \qquad (1)$$

where $L_2$ = 3 μm is the length of the coupled waveguide in the Directional Coupler 2. Figure 2c shows the normalized output power in the first waveguide as a function of the SW frequency $f$ in the frequency range from 2.28 GHz to 2.65 GHz. The result of numerical simulations in the linear regime is shown with the blue symbols and the analytic calculation with the blue solid lines (see Methods for details). One can clearly see, that the output power $P_{1out}$ strongly depends on the SW frequency. This is due to the strong dependence of the coupling length $L_{C2}$ on the SW wavenumber [36,37]. The coupling length, consequently, defines the energy distribution between the output waveguides for a given length of the coupled waveguides. The small mismatch between simulations and theory in the region below 2.3 GHz is mainly caused by the damping, which is not taken into account in the theory, and by the large sensitivity of the coupling coefficient to the dispersion of the antisymmetric mode, which is practically flat in this region.

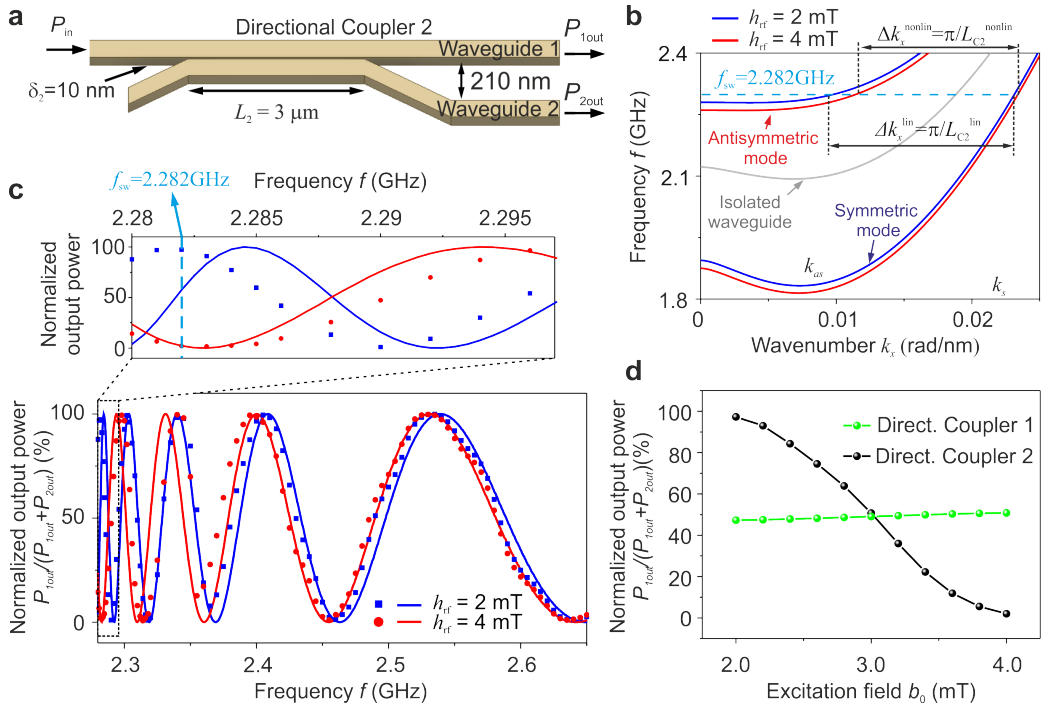

**Fig. 2 Modeling and characteristics of the Directional Coupler 2**. **a** Schematic of the directional coupler. **b** Analytically-calculated dispersion curves for the coupled waveguides for small (blue lines) and large (red lines) excitation fields $h_{rf}$. The change in the coupling length $L_{C2}$ is clearly visible that is associated with the decrease in the SW frequency with the increase in its amplitude. **c** Normalized output power in the first waveguide $P_{1out}/(P_{1out}+P_{2out})$ as a function of frequencies for different excitation field $h_{rf}$ (symbols - simulations, lines – analytic theory). A zoom of the region marked with the dashed rectangle is shown in the top panel. **d** Simulated normalized output power $P_{1out}$ as a function of the excitation field $b_0$ for a fixed frequency of $f$ = 2.282 GHz for Directional Coupler 1 and 2.

When the input SW power increases, nonlinear effects start to play an important role. In the range of relatively weak nonlinearity, the main impact is produced by the nonlinear frequency shift of the SW [16,38,43-45]. Since the dipolar coupling between waveguides is much smaller than the internal forces, i.e., the splitting of the collective SW modes is much lesser than the SW frequency, we can neglect a nonlinear correction to the coupling strength. In this case, the power-dependent SW dispersion of the collective modes of coupled waveguides can be calculated as [46]:

$$f_{s,as}^{(nl)}(k_x, a_{k_x}) = f_{s,as}^{(0)}(k_x) + T_{k_x} |a_{k_x}|^2 \qquad (2)$$

where $f_{s,as}^{(0)}(k_x)$ are the dispersion relations of symmetric and antisymmetric modes of the coupled waveguides in the linear regime, $a_{k_x}$ is the dimensionless SW amplitude and $T_{k_x}$ the nonlinear frequency shift of SWs in an isolated waveguide. This frequency shift is mainly due to the decrease in the effective magnetization of a magnetic material with the increase in the magnetization precession angle [44,45]. All these characteristics are described in more detail in the Methods Section. For the further calculations, we used the numerically found relation between the excitation field and the amplitude of the dynamic magnetization with both in units of mT. In in-plane magnetized structures the nonlinear shift is negative [46] and, thus, according to Eq. (2), the SW dispersion shifts down with an increase in SW amplitude – see red lines in Fig. 2b. Thus, a fixed SW frequency of 2.282 GHz becomes correspondent to different SW wavenumbers what changes the parameter $\Delta k_x = \pi/L_{C2}$ from $\Delta k_x^{\text{lin}}$ to $\Delta k_x^{\text{nonlin}}$ (see red dots and lines in Fig. 2) with an increase in the excitation field from $b_0 = 2$ mT to 4 mT. Consequently, the coupling length $L_{C2}$ of the directional coupler also changes.

Using the Taylor expansion of the frequency dependence of the coupling length, the power dependence of the output of Directional Coupler 2 can be found:

$$\frac{P_{1\text{out}}}{P_{1\text{out}} + P_{2\text{out}}} = \cos^2\left(\frac{\pi L_2}{2L_{C2}^{\text{lin}}} - \frac{L_2}{L_{C2}^{\text{lin}}}\frac{\pi}{2L_{C2}^{\text{lin}}}\frac{\partial L_{C2}}{\partial f}T_{k_x}|a_{k_x}|^2\right) \qquad (3)$$

The power-independent term is proportional to the ratio of the directional coupler length to the coupling length in the linear regime $L_2/L_{C2}^{\text{lin}}$. The output power $P_{1\text{out}}$ periodically changes with a change in the coupling length and is maximal for the cases $L_2/L_{C2}^{lin}$ = 0, 2, 4, … (see Fig. 2c). Simultaneously, as it is seen from Eq. (3), the sensitivity to the nonlinear effect increases with the increase in the ratio $L_2/L_{C2}^{lin}$. Therefore, the longer the directional coupler is and the more coupling lengths it spans, the higher is the nonlinear phase accumulation. This is the reason why the Directional Coupler 2 in our half-adder design is long and features a strong coupling provided by the small gap between the waveguides of only 10 nm. It has a length of $L_2 = 14L_{C2}^{lin}$ and it is very sensitive to the increase in the SW amplitude passing through it. As a result, a complete energy transfer from Output 1 to Output 2 is observed in the micromagnetic simulations if the SW amplitude is increased by a factor of two ($L_2 = 13L_{C2}^{nonlin}$) – see black line in Fig. 2d. The normalized output SW power in the first waveguide decreases from 97.3% at $b_0$ = 2 mT to 2.0% at $b_0$ = 4 mT. Due to this nonlinear switching effect, the Directional Coupler 2 performs a combined AND and XOR logic function, as will be described in the following.

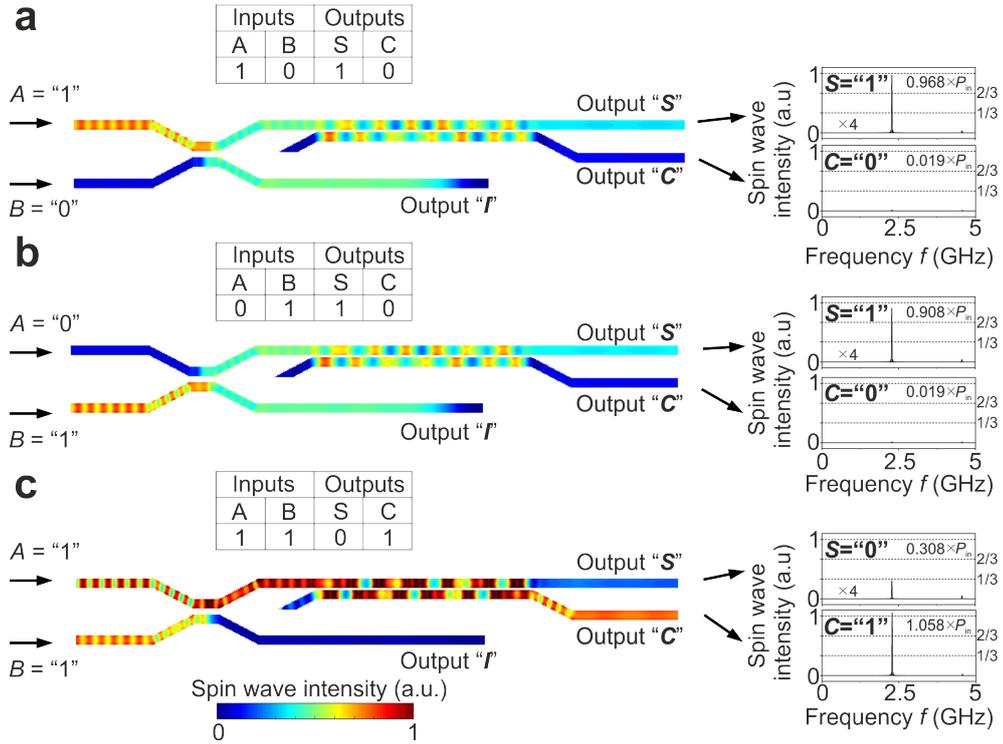

**Fig. 3 Operational principle of the magnonic half-adder.** The SW intensity distributions for different inputs combinations: **a** "*A*" = "1", "*B*" = "0", **b** "*A*" = "0", "*B*" = "1" and **c** "*A*" = "1", "*B*" = "1". The truth tables for each inputs combinations are shown on the top of the structure. The normalized output SW intensities in the outputs are shown on the right side. All the outputs "*S*" are multiplied by a factor of four.

At the same time, the first Directional Coupler 1 in the half-adder design should remain in the linear regime and its coupling length should be independent on the SW amplitude. This is achieved by its smaller length of 370 nm as well as by a decreased strength of the coupling via an increased spacing between the waveguides of 50 nm. As a result, Directional Coupler 1 spans only half of the coupling length $L_1 = 0.5 L_{C1}^{lin}$. Thus, it requires much higher SW amplitudes to show a power dependence and in our working range of SW amplitudes, its output remains around 50% and is independent on the excitation field – see green symbols in Fig. 2d. Hence, in contrast to Directional Coupler 2, Directional Coupler 1 works as a stable energy splitter, combiner and bridge for the two input signals.

**Operational principle of the magnonic half-adder.** The operational principle of the half-adder is shown in Fig. 3. Binary data is coded into the SW amplitude, namely, in the ideal case, a spin wave of a given amplitude (e.g., $M_z/M_s = 0.057$) corresponds to the logic state "1" while zero SW amplitude corresponds to "0". In the following, we normalize all output SW intensity to the input SW intensity. In the more realistic cases considered below, we utilize an approach from CMOS: a normalized SW intensity below 1/3 is considered to be logic "0" and above 2/3, it is logic "1".

The operational principle of the half-adder is as follows. Let us first consider the case of logic inputs "*A*" = "1" and "*B*" = "0" – see Fig. 3a. In this case, the SW injected into the Input "*A*" is split into two equal parts by the Directional Coupler 1. One of them is directly guided to the Directional Coupler 2 by the upper conduit. The SW intensity is chosen in such a way that Directional Coupler 2 remains in the linear regime ($L_{C2}^{lin} \approx 214$ nm $\approx 14/L_2$) and after initial oscillations, the SW is guided into the Output "*S*" as shown in the figure. Only about 1.9% of the SW energy goes into the Output "*C*". This corresponds to the logic Outputs "*S*" = "1" and "*C*" = "0". If a SW is injected in Input "*B*" only, this corresponds to the logic inputs *A*" = "0" and "*B*" = "1" – see Fig. 3b. The situation in this case is quite similar to the previous one: The SW intensity is split into two parts by the Directional Coupler 1, one of which is guided to the output "*S*" via the Directional Coupler 2 and, thus, again "*S*" = "1", "*C*" = "0". The situation is different for the input logics states "*A*" = "1" and "*B*" = "1" – see Fig. 3c. It is assumed that the phase of the SW injected into the Input "*B*" is shifted by π/2 with respect to the one in the Input "*A*" in order to compensate the -π/2 phase shift caused by the Directional Coupler 1. Since the spin-wave wavelength is fixed as well as the required phase shift, in practice, the phase shifter can be realized by varying, for instance, the length of the magnonic conduits between two half-adders in series, by varying the waveguide width or by creating inhomogeneities in the biasing magnetic field. As a result, the excitation of SWs in both inputs results in their constructive interference and in a twice larger SW amplitude arriving at Directional Coupler 2. As was discussed above, this increase in the SW intensity switches the coupler to the nonlinear regime ($L_{C2}^{nonlin} \approx 230$ nm $\approx 13/L_2$) and the spin wave is guided to the output "*C*". This corresponds to the logic Outputs "*S''* = "0" and "*C''* = "1" (see Fig. 3c) and, thus, the whole truth-table of the half-adder is realized. The combination of two directional couplers preforms the AND and XOR logic functions due to the phenomenon of nonlinear SW frequency shift.

Please note that the all-magnon circuit concept [15] requires that the signal from the output of a magnonics gate can be directly guided into the input of the next one. In order to satisfy this condition, the spin-wave intensity at the Outputs "*S''* still has to be amplified by a factor of four due to the energy splitting in the Directional Coupler 1 and due to parasitic reflections and spin-wave damping in the waveguides. The output signals "*S*" shown in Fig. 3 are artificially multiplied by 4. The most promising realization of such an amplifier is based on the utilization of parametric pumping [47,48]. In contrast, the SW amplitude in the output "C" is either vanishingly small or is approximately equal to the input SW amplitude as a consequence of constructive interference. Thus, no amplifier is required for the "Carry" output of the half-adder. In general, the idea presented here and the concept of the half-adder is applicable for any magnetic material. Nevertheless, the requirement that the device length $L_{de}$ is smaller than the spin-wave decay length should be satisfied. Thus, the materials with large values of spin-wave lifetimes and group velocities, as well as directional coupler with a strong coupling between the waveguides are preferable.

**Discussion**

In the final part of this paper, we would like to discuss potential power consumption, scalability, and delay time of the presented magnonic half-adder and also analyze the energy consumption of different spin-wave amplifiers.

**Energy consumption**: For the estimation of energy consumption in the magnonic system (neglecting transducers), the minimal energy consumption can be express as (see Supplementary Materials for the details of the derivation):

$$E = \frac{20\pi}{3} \frac{M_s}{\gamma} \frac{v_{gr} f S}{T_{k_x}} \tag{4}$$

where $v_{gr} = 2\pi \frac{\partial f}{\partial k}$ is the SW group velocity, $S$ is the cross-section of the waveguide. As one can see, the energy consumption is independent of the characteristics of SW couplers and SW amplitude. Note that the nonlinear frequency shift $T_{k_x}$ is of the order of the SW frequency $f$ ($T_{k_x} \propto f$), especially in the exchange-dominated region. The conclusion has arrived that the feasible way to reduce the energy consumption is the decrease of waveguide cross-section $S$. Another alternative is searching for specific points or mechanisms with anomalously high nonlinearity. The use of exchange spin waves will not decrease the energy consumption. Instead, the SW group velocity increases in the exchange region which results in an increase of the energy consumption. It should be noted that the relation of the Eq. (4) is universal and takes place in other realizations of magnonic half-adders which are based on the nonlinear shift. For the other designs, the only change is the pre-factor $20\pi/3$.

**Scalability and delay time**: The width of the device can be estimated by:

$$w_{de} = 2w + 4 \times 5h \tag{5}$$

where $w$ is the width of waveguide and $h$ is the thickness of waveguide. This equation accounts for the minimal distance between all waveguides and neighboring devices at $5h$ to make dipolar interaction relatively weak. The gaps between different logic gates are taken into account in this width.

The length of the device is given by

$$L_{de} = (N + 0.5) L_C + 4 \frac{5h}{\sin\varphi} \tag{6}$$

where $\varphi$ is the angle of the bent waveguide, $L_C$ is the coupling length, $N = L_2/L_C$ is the ratio between the coupled length of Directional Coupler 2 and the coupling length. The minimal $N$ can be estimated from the condition that Directional Coupler 1, working at half the coupling length, does not significantly change its characteristics at power, which is sufficient to switch the Directional Coupler 2. Simple calculations yield that the change of Directional Coupler 1 transmission is given by $\cos^2(\pi(N-1)/(4N))$, while in the linear regime transmission rate is equal to 1/2. This gives the restriction $N_{min}$ = 6. The area of the magnonic half-adder is $Area = w_{de}L_{de}$. The processing delay is $\tau_d = L_{de}/v_{gr}$.

Figure 4 presents results for the half-adder with the dimensions which have been mentioned above. The only difference is that the minimal $N$ = 6 is used instead of $N$ = 14 used in the simulations. The total energy consumption and delay time are very close to the estimations. The area of half-adder is slightly larger than the estimation which can be optimized by decreasing the gaps between the two input waveguides and the length of the Directional Coupler 2.

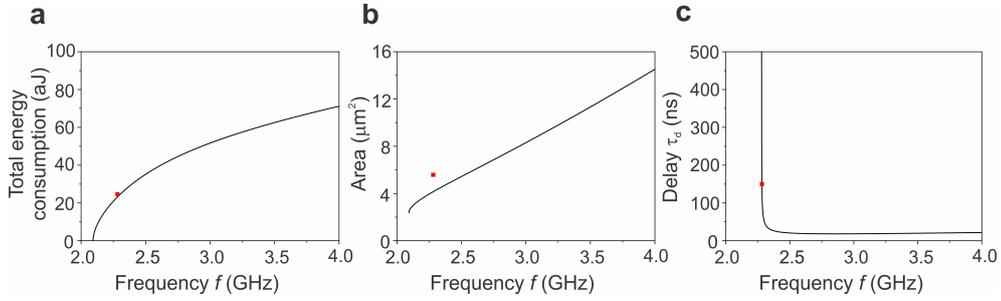

**Fig. 4 Estimation of energy consumption, scalability and delay time of the magnonic half-adder.** The energy consumption (a), scalability (b), and delay time (c) as a function of SW frequency. The black lines are obtained from Eqs. (4)-(6). The red dots are extracted from simulations.

The summary of these parameters is shown in Table 1. Parameters for different devices are estimated: The first one is the device that was simulated and discussed above. The width and thickness of the waveguides are equal to 100 nm and 30 nm which can be reliably fabricated according to the recent achievements [41, 42]. The second device is an estimation performed for a device with $w$ = 30 nm, $h$ = 10 nm and minimal gap $\delta$ = 10 nm. It has to be mentioned that the second device does not constitute a fundamental limit but is merely an estimation based on the current state of the art of fabrication technology. A further improvement in all characteristics is potentially achievable by introducing another coupling mechanism instead of dipolar coupling.

| Parameters | YIG * (100 nm) | YIG † (30 nm) | CMOS ‡ (7 nm) |
|---|---|---|---|
| Area (μm²) | 5.58 | 1.016 | 1.024 |
| Delay time (ns) | 150 | 18 | $6\times10^{-2}$ |
| Total energy consumption without amplification (aJ) | 24.6 | 1.96 | 35.3 |
| SW frequency (GHz) | 2.282 | 2.29 | - |
| SW wavelength (nm) | 340 | 510 | - |
| SW group velocity (m/s) | 25 | 137 | - |

| Type of amplifier | Energy consumption |
|---|---|
| Electric current based parametric pumping [47] | $10^5$ (aJ/operation) |
| Voltage controlled magnetic anisotropy parametric pumping [48,49] | 3 (aJ/operation) |

**Table 1.** Characteristics of the half-adder investigated in the paper, estimated characteristics of a device miniaturized down to 30 nm, the calculated parameter of the current state of the art of 7 nm CMOS. * The column is extracted from micromagnetic simulation. † The column is estimated using the Eqs. (4) - (6). ‡ The column is calculated using Cadence Genus by Sorin D. Cotofana for 7 nm CMOS technology.

According to the table, the area of the simulated 100 nm feature size half-adder is 5.58 μm² (the spaces between the neighboring logic gates are also included) and is, thus, only a few times larger than a corresponding 7 nm feature size CMOS device. In contrast to a CMOS realization, the magnonic half-adder core part (without amplifier) consists of only three nano-wires made of the material and of only one planar layer. This drastically simplifies its fabrication and decreases its potential costs. The area can be readily decreased down to 1.016 μm² for the second device. In addition, it should be noted that the largest part of the half-adder is given by the Directional Coupler 2, which could potentially be further decreased via the utilization of other, stronger coupling mechanisms between the SW waveguides like exchange instead of dipolar coupling. To achieve this, the air gap between the coupled waveguides should be filled with another magnetic material.

Operational frequency is an important requirement. In the presented half-adder, the delay time is defined by the whole length of the device with respect to the SW group velocity. In our design, the SW propagation time from input to output is about 150 ns. According to Table 1, the calculation time can be reduced to 18 ns in the second device. This value is still much larger than the 60 ps delay time obtained for 7 nm

CMOS and assumes that magnon logic would be more suitable for the slow but low-energy applications.

Small energy consumption of computing systems is probably the most crucial requirement taking into account the constantly increasing amount of information that has to be processed. In our simulations, we record the total energy of the device as a function of simulation time. The energy injected into the device per nanosecond is equal to $4.1 \times 10^{-20}$ J/ns for the input combinations "$A$" = 1, and "$B$" = 1. Please note that only the energy propagating along the positive direction is taken into account. For the 300 ns pulse duration the energy consumption is, thus, 12.3 aJ. For all the operations, the total energy consumption is 24.6 aJ. This is similar to current CMOS values (35.3 aJ) which is calculated using Cadence Genus (see Method). It should be especially highlighted that the energy consumption of the miniaturized device is remarkably low which is around 1.96 aJ. At the same time, we have to underline that this energy consumption is related to the energy within the magnonic domain only and it does not take into account the energy consumed during the transfer of signals between the elements and the energy required for the SW amplification.

Finally, the operation of the proposed all-magnon half-adder requires the utilization of the amplifiers at the output "S" in order to bring its amplitude to the level of the input signal. The energy consumption of the amplifier should be added to the consumption of the half-adder itself. Currently, from all the variety of spin-wave amplifiers, the following two approaches seem to represent the main interest. The first one is the parallel parametric pumping which converts microwave photons into magnons at half of the microwave frequency [47]. Such amplifier was simulated (see details in the Supplementary Materials) and its energy consumption appeared to be around $10^5$ aJ (see Table 1), which can be further decreased one or two orders of magnitude by optimizing the parameters of the pumping antenna and the half-adder. In general, this energy is much higher than the energy used to process data within the magnonic domain. The second approach relies on the usage of the voltage control magnetic anisotropy parametric pumping that has been reported recently [48]. This approach uses rf electric rather than the magnetic (and associated electric current) field and, therefore, allows for the drastic decrease in the energy consumption down to ~ 3 aJ per device (was estimated for a nm-thick CoFeB film - see the Supplementary Materials for details). The experimental realization of such a device was reported recently [49]. The same approach should also be applicable for YIG if to place an additional layer of piezoelectric on top [50].

In conclusion, we have proposed and numerically tested a magnonic half-adder that constitutes of two logic gates and is potentially suited for further integration with other logic gates relying on the same concept. In this integrated magnonic circuit, the magnons are controlled by magnons themselves without any conversion to the electric domain which ensures low energy consumption. The half-adder consists of two directional couplers, one of which acts as a power splitter and another which acts as a nonlinear switch. The operational principle of the latter is based on the nonlinear shift of spin-wave frequency with an increase in spin-wave amplitude. An analytic theory is developed to describe this phenomenon and is verified numerically as well as the whole functionality of the half-adder.

In comparison to CMOS technology, the half-adder consists of three magnetic nano-wires with one amplifier placed on top of the output "$S$" and substitutes 14 transistors in electronics. The magnonic half-adder, although being based on 100 nm technology, has characteristics comparable to a 7 nm CMOS half-adder. At the same time, the magnonic half-adder has large potential for miniaturization and further improvement.

**Methods**

**Spin-wave dispersion calculation in the linear region.** The details of the calculations have been fully described in our previous paper [36]. The SW dispersion curve can be obtained by solving the Landau-Lifshitz equation of magnetization dynamics in the linear approximation and neglecting the damping term. The SW dispersion relation in the isolated waveguide is given by [36]:

$$f_0(k_x) = \frac{1}{2\pi}\sqrt{\Omega^{yy}\Omega^{zz}} = \frac{1}{2\pi}\sqrt{\left(\omega_H + \omega_M\left(\lambda^2 k_x^2 + F_{k_x}^{yy}(0)\right)\right)\left(\omega_H + \omega_M\left(\lambda^2 k_x^2 + F_{k_x}^{zz}(0)\right)\right)} \quad (7)$$

The dispersion relation for the two coupled waveguides (two modes) is

$$f_{s,as}(k_x) = \frac{1}{2\pi}\sqrt{\left(\Omega^{yy} \pm \omega_M F_{k_x}^{yy}(d)\right)\left(\Omega^{zz} \pm \omega_M F_{k_x}^{zz}(d)\right)} \quad (8)$$

where $\Omega^{ii} = \omega_H + \omega_M\left(\lambda^2 k_x^2 + F_{k_x}^{ii}(d)\right)$, $i = y, z$, $\omega_H = \gamma B_{ext}$, $\omega_M = \gamma\mu_0 M_s$, $M_s$ is the saturation magnetization, $\gamma$ is the gyromagnetic ratio, $\mu_0$ is the vacuum permeability, $\lambda = \sqrt{2A/(\mu_0 M_s^2)}$ is the exchange length, $A$ is the exchange constant, $d = w + \delta$ is the distance between the centers of the two waveguides, $\delta$ is the gap between the two waveguides with width $w$. The coordinate system used is shown in Fig. 1c: The $x$-axis is directed along the waveguides, the $z$-axis is out-of-plane. The tensor $\hat{\mathbf{F}}_{k_x}$ describes the dynamical magneto-dipolar interaction [51,52]:

$$\hat{\mathbf{F}}_{k_x}(d) = \frac{1}{2\pi}\int \hat{\mathbf{N}}_k e^{ik_y d} dk_y \tag{9}$$

$$\hat{\mathbf{N}}_k = \frac{|\sigma_k|^2}{\tilde{w}}\begin{pmatrix} \frac{k_x^2}{k^2}f(kh) & \frac{k_x k_y}{k^2}f(kh) & 0 \\ \frac{k_x k_y}{k^2}f(kh) & \frac{k_y^2}{k^2}f(kh) & 0 \\ 0 & 0 & 1-f(kh) \end{pmatrix} \tag{10}$$

where $f(kh) = 1-(1-\exp(-kh))/(kh)$, $k = \sqrt{k_x^2 + k_y^2}$, $h$ is the thickness, $\sigma_k$ is the Fourier transform of the SW profile across the width of the waveguide, and $\tilde{w}$ is the normalized constant of the mode profile. In this case, the spins are fully unpinned at the edge of the waveguides [41]. The Fourier transform is, then, described by the function $\sigma_k = w\mathrm{sinc}(k_y w/2)$ and $\tilde{w} = w$.

**Nonlinear frequency shift.** The nonlinear shift coefficient $T_{k_x}$ in the isolated waveguide can be calculated using the framework of [53] and by assuming a uniform mode profile across the waveguide thickness and width. Accounting for the negligible static demagnetization of a waveguide along its length, $F_0^{xx} = 0$, the nonlinear shift coefficient becomes equal to [46]:

$$T_{k_x} = \left(\left(\omega_H - A_{k_x}\right) + \frac{B_{k_x}^2}{2\omega_0^2}\left(\omega_M\left(4\lambda^2 k_x^2 + F_{2k_x}^{xx}(0)\right) + 3\omega_H\right)\right)\bigg/2\pi, \tag{11}$$

where

$$A_{k_x} = \omega_H + \frac{\omega_M}{2}\left(2\lambda^2 k_x^2 + F_{k_x}^{yy}(0) + F_{k_x}^{zz}(0)\right), \tag{12}$$

$$B_{k_x} = \frac{\omega_M}{2}\left(F_{k_x}^{yy}(0) - F_{k_x}^{zz}(0)\right). \tag{13}$$

The relation between the dynamic magnetization component and the canonical SW amplitude $a_{k_x}$ is given by:

$$M_z = M_s a_{k_x}\sqrt{2-|a_{k_x}|^2}\left(u_{k_x} - v_{k_x}\right), \tag{14}$$

with

$$u_{k_x} = \sqrt{\frac{A_{k_x} + \omega_0}{2\omega_0}} \text{ and } v_{k_x} = -\text{sign}\left[B_{k_x}\right]\sqrt{\frac{A_{k_x} - \omega_0}{2\omega_0}}. \quad (15)$$

**Micromagnetic simulations**. The micromagnetic simulations were performed by the GPU-accelerated MuMax3 [54] code. The simulated structure of the magnonic half-adder is shown in Fig. 1c. The parameters of nanometer thick YIG are obtained from experiment and are as follows [11]: saturation magnetization $M_s = 1.4 \times 10^5$ A/m, exchange constant $A = 3.5$ pJ/m, and Gilbert damping $\alpha = 2 \times 10^{-4}$. The damping at the ends of the simulated structure and the high damping absorber is set to exponentially increase to 0.5 to prevent SW reflection [55]. The high damping region could be realized in the experiment by putting another magnetic material or metal on top of YIG to enhance the damping or it can just correspond to waves guided into further parts of the magnonic network. No external bias field is applied. The static magnetization orients itself parallel to the waveguides spontaneously due to the strong shape anisotropy in the nanoscale waveguides. The mesh was set to $10 \times 10 \times 30$ nm$^3$. To excite propagating spin waves, a sinusoidal magnetic field $b = b_0\sin(2\pi ft)$ is applied over an area of 100 nm in length, with a varying oscillation amplitude $b_0$ and the microwave frequency $f$. $M_z(x,y,t)$ of each cell was collected over a period of 300 ns which is long enough to reach the steady state. The fluctuation $m_z(x,y,t)$ were calculated for all cells via $m_z(x,y,t) = M_z(x,y,t) - M_z(x,y,0)$, where $M_z(x,y,0)$ corresponds to the ground state. The SW spectra of the output signals are calculated by performing a fast Fourier transformation from 250 ns to 300 ns which corresponds to the steady state. We would like to mention that all these simulations were performed for the defect-free waveguides and without taking into account temperature. The influences of the edge roughness, trapezoidal cross-sections of the waveguides, and temperature can be ignored due to their smallness as it has been shown in our previous studies [36,41].

**Calculation of energy consumption of 7 CMOS half-adder.** We considered a 7nm HA standard cell afferent to the typical processor corner (room temperature, 0.7V power supply) and evaluated its power consumption using Cadence Genus. To this end, we set an inverter standard cell as driver and a capacitance of 2.5fF as output load, and assume for the nets a 50% probability of logic "1" and a toggle rate of 0.02 per ns.

Simulation results indicate a total power consumption of 587.994nW out of which the dynamic component (divided into nets power and internal power, which account for 87.7% and 13.3% of the dynamic power, respectively) dominates the less than 1nW leakage component.

**Acknowledgements:**

The authors thank Burkard Hillebrands for support and valuable discussions. This research has been supported by ERC Starting Grant 678309 MagnonCircuits and by the DFG through the Collaborative Research Center SFB/TRR-173 "Spin+X" (projects B01) and through the Project DU 1427/2-1. R. V. acknowledges support from the Ministry of Education and Science of Ukraine, Project 0118U004007.

**Author contributions:**

Q.W proposed the magnonic half-adder design, performed micromagnetic simulation jointly with T. B. and P. P., and wrote the first version of the manuscript. T. B. and A. V. C. devised and planned the project. R. V. developed the analytical theory and estimated the energy consumption of the half-adder. F. C., C. A. and S. D. C. performed the benchmarking and calculated the parameters of 7 nm CMOS half-adder. P. P and A. V. C led the project. All authors discussed the results and contributed to writing the manuscript.

# Supplementary Material
# Integrated magnonic half-adder

Qi Wang[1], Roman Verba[2], Thomas Brächer[1], Florin Ciubotaru[3], Christoph Adelmann[3], Sorin D. Cotofana[4], Philipp Pirro[1], and Andrii V. Chumak[1,5,†]

[1] Faculty of Physics and Research Center OPTIMAS, University of Kaiserslautern, Kaiserslautern 67663, Germany

[2] Institute of Magnetism, Kyiv 03142, Ukraine

[3] Imec, Leuven, 3001 Belgium

[4] Department of Quantum and Computer Engineering, Delft University of Technology, Delft, 2600 The Netherlands

[5] Faculty of Physics, University of Vienna, Boltzmanngasse 5, A-1090 Vienna, Austria

In the supplementary material, we first discuss the details of the estimation of the energy consumption for magnonic half-adder in Section S1. The modified device operates as a combination of a half-adder with a fan-out gate is discussed in Section S2. In Section S3, we analyzed three different spin-wave amplifiers and estimated the energy consumption of two of them.

## S1. Energy consumption for magnonic half-adder

For the estimation of energy consumption in the magnonic system (neglecting transducers), we can simply calculate the energy of the SW pulse. The energy of SW pulse is given by

$$E = \frac{M_s V}{\gamma} \omega_k |c_k|^2 \qquad (1)$$

where SW amplitude $c_k$ is defined so that SW norm $A_k = 1$ (which is the same as common $u$-$v$ transformation), and $V$ is the volume of SW pulse. Naturally, $V = S v_{gr} \tau$, with $S$ being the cross-section of the waveguide, $v_{gr}$ – SW group velocity and $\tau$ – pulse duration.

The SW amplitude is determined from the condition that it should be enough to switch nonlinear directional coupler. From Eq. (3) of the paper, we have the following condition

---

$$\frac{L}{L_c}\frac{\pi}{2L_c}\frac{\partial L_c}{\partial f}T_k|c_k|^2 = \frac{\pi}{2} \tag{2}$$

For simplicity, we neglect the sign of $T_k$ and $\partial L_c/\partial f$ – they are not important. Then, using the definition of the coupling length via the coupling frequency $\Omega$, and setting $L/L_c=N$, $N = 0, 2, 4,\ldots$ (i.e., in the linear regime we work on the $N/2$ - the transmission maximum), we get

$$\frac{N}{\Omega}\frac{\partial \Omega}{\partial f}T_k|c_k|^2 = 1 \tag{3}$$

The duration of the SW pulse is determined by the maximal frequency width of the pulse $\Delta\omega$, $\tau = 2\pi/\Delta\omega$, which should be small enough for stable operation of both linear and nonlinear couplers. In the case of the nonlinear coupler, which works near the maximum or minimum of transmission,

$$\cos\frac{\Omega L}{v_{gr}} \approx 1 - \frac{1}{2}\left(\frac{L}{v_{gr}}\right)^2 \Delta\Omega^2 \tag{4}$$

where $\Delta\Omega = (\partial\Omega/\partial\omega)\Delta\omega$ and the last term should be $\ll 1$. Using the convention that $x \ll 1$ means $x \leq 0.1$, we get

$$\frac{\Delta\Omega}{\Omega} < \frac{0.3}{N} \tag{5}$$

Performing the same calculations for a linear coupler (assuming its length $L=L_c/2$), we get $\Delta\Omega/\Omega < 0.2$. In fact, the condition for the nonlinear case is more severe, as $N \geq 2$. More rigorously, the minimal size of the nonlinear coupler $N$ can be estimated from the condition that the linear coupler, working at half the coupling length, does not significantly change its characteristics at the power which is enough to the switch nonlinear coupler. Simple calculations yield that the change of linear coupler transmission is given by $\cos^2(\pi(N-1)/(4N))$, while the linear regime transmission rate is equal to 1/2. This gives the restriction $N_{\min} = 6$ (assuming that linear and nonlinear coupler have the same gap).

Combining Eqs. (5, 3, 1), we get the final expression for the minimal energy consumption (Eq. 4 in the paper)

$$E = \frac{20\pi}{3}\frac{M_s}{\gamma}\frac{v_{gr}fS}{T_k} \tag{6}$$

## S2 Modified half-adder with fan-out gate

In this manuscript, the Output "I" features high damping to avoid the reflection at the end of the waveguide. However, the SW amplitude in this waveguide satisfies the XOR logic gate – see truth table Table S1. In the further, it could also be used for logical operation.

| Inputs | | Output |
|---|---|---|
| A | B | I |
| 0 | 0 | 0 |
| 1 | 0 | 1 |
| 0 | 1 | 1 |
| 1 | 1 | 0 |

Table S1. Truth of Output "I"

Furthermore, the Output "I" could also be used to perform the half-adder operation. In order to do that, an additional coupler was introduced to the Output "I" as shown in Fig. S1(a)-(c). For this, another waveguide is closely placed to the Output "I" and the phase difference between the inputs "A" and "B" has been changed from $\pi/2$ to 0. The operation principle is identical to the original half-adder for the inputs combinations "A" = "1", "B" = "0" and "A" = "0", "B" = "1" as shown in Fig. S1 (a)-(b). If the inputs are "A" = "1", "B" = "1", the excitation of spin waves in both inputs results in the transfer of spin waves of the same intensities to both directional couplers.

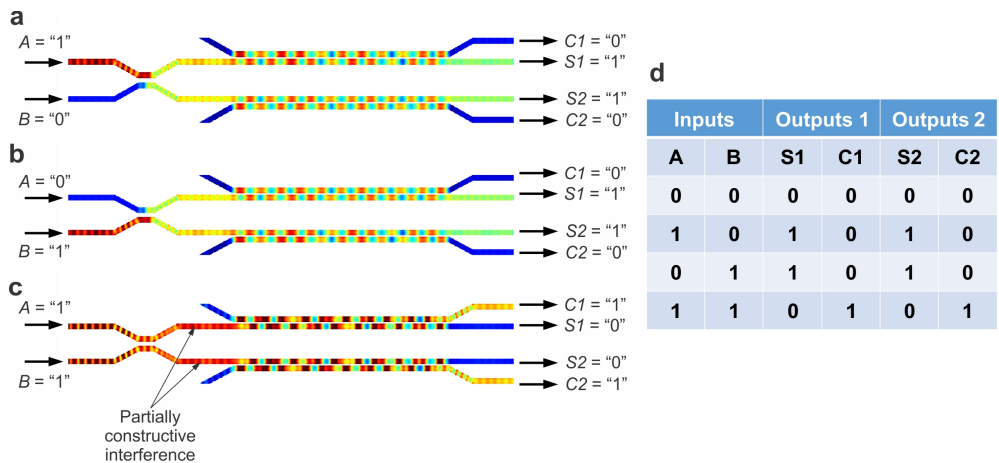

Fig. S1 (a-c) The operational principle and (d) truth table of two half-adders with shared inputs (corresponds to the half-adder with added fan-out logic gate).

In this case, the SW intensity reached the nonlinear directional coupler is twice comparing to the single input (but not four times larger like in the original half-adder design). Nevertheless, even the increase in the SW intensity twice between the logic states allowed for the triggering of the nonlinear switching phenomenon (after adapting working frequency and the coupled length of the device) - see Fig. S1(c). This device operates as a combination of half-adder with fan-out gate and each logic output is doubled.

**S3. Energy consumption of parametric amplification**

The amplifier is a very important point for the whole all-magnon data processing concept since the amplitude of the output signal should always be equal to the input one. The energy consumption of the amplifier might be crucial. To find the answer which amplifier is the most suitable, we analyzed three different approaches: Spin-Orbit Torque-based approach, electric current-based parametric pumping, and Voltage Control Magnetic Anisotropy (VCMA) electric field-based parametric pumping.

1) The first one is based on the usage of, so-called Spin-Orbit Torque (the combination of spin Hall effect (SHE) with spin transfer torque (STT)) [e.g., described in the review [1]. Unfortunately, this approach does not seem to be suitable since the amplification of the propagating spin waves was not demonstrated so far. Even the result reported in Ref. [2] hardly can be used to amplify the propagation spin waves since the final smallest damping parameter in bilayer YIG/Pt with the current was comparable to the original damping of pure single layer YIG. The spin-orbit torque can only compensate for the damping caused by spin pumping due to the introduction of the Pt layer. To sum up, the output spin-wave intensity is always lower than the input in this system. Moreover, this approach requires the applications of electric currents of high densities and, therefore, hardly can ensure small energy consumption.

2) The second way is a classical parallel parametric pumping process in which the pumping rf magnetic Oersted field is created by sending of rf signal through a metallic strip [3]. This is the most well-studied approach in magnonics. In order to find the energy consumption of parametric amplification, we perform simulations in a single waveguide with a pumping antenna. Figure S2(a) shows the schematic picture of the parametric pumping region where an antenna (width: $w_a$ = 500 nm, thickness: $t_a$ = 300 nm) is placed on top of the YIG waveguide.

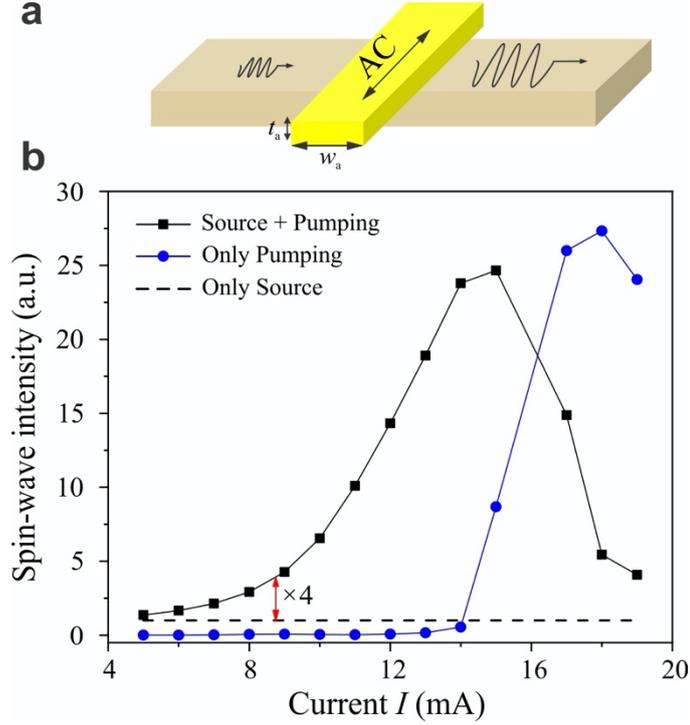

*Fig. S2 (a)A schematic picture of the parametric amplifier. (b) Output spin-wave intensity as a function of pumping current for different conditions.*

In the simulations, the pumping frequency (2$f$) is twice the frequency of the initial spin wave. The spin-wave intensity at the frequency of $f$ is extracted after the pumping region and is plotted as a function of pumping current. The blue dotted line shows the spin-wave intensity when only the pumping current is applied, where a clear threshold is observed. The intensity of the initial spin wave is marked by the dashed line. The black dotted line presents the total energy of the output spin-wave after the amplification when the pumping current and initial spin wave are applied simultaneously. The gap between the dashed line and the black dotted line indicates the amplification due to the parallel pumping. It is worth to mention that the parametric pumping starts to amplify the initial spin wave before the it reaches the threshold of the parametric generation. This means that the parametric pumping will not directly excite parasitic spin waves in the magnonic circuits. The spin-wave intensity drops down in the high pumping current region due to the four magnon scattering.

In order to obtain a 4 times amplification of SW intensity, which is needed for the magnonic half-adder, the pumping current should be around 8.8 mA as shown in Fig. S2(b). For each output "S", the length of the pumping antenna is at least $L_a$=100 nm (the width of waveguide). Then, the resistance of the pumping antenna can be calculated using $R=\rho L_a/(w_a t_a)$=0.016 Ohm ($\rho = 2.44\times10^{-8}$ Ohm/m is the resistivity of gold at room temperature). The pumping time is 300 ns. Thus, the energy consumption of amplification is around $10^5$ aJ, which can be further decreased one or two orders of magnitude by optimizing the size of the pumping antenna and the half-adder. In general, this energy consumption is much higher than the energy used to process data within magnonic domain.

3) The most suitable approach appeared to be the VCMA parametric pumping that has been reported in Ref. [4]. In this case, the parametric process is similar to the previous one but instead of the magnetic rf field (which requires large currents), an electric rf field is used. This allows for the drastic decrease in the energy consumption of the amplification down to 3 aJ as was estimated for the example of a nm-thick CoFeB waveguide. The functionality of such kind of parametric pumping was recently reported experimentally in [5].

The detail of estimation is given below. For VCMA parametric pumping, the electric field required is equal to:

$$E = \frac{hM_s b_{th}}{2\beta} \tag{7}$$

with $\beta$ – magnetoelectric coefficient and $h$ – waveguide thickness. Losses consists of dielectric losses and Ohmic ones,

$$W = \left( E^2 d\omega\varepsilon\varepsilon_0 wL \tan\delta + \frac{E^2 d^2}{R} \right)\tau \tag{8}$$

where $d$ – dielectric thickness, $\varepsilon$ – dielectric permittivity, $\tan\delta$ – tangents of dielectric losses, $R$ – tunnel resistance. Both terms depend on the thickness $d$ and the total losses are minimal at certain optimal $d$.

In the estimation, we use $L = 100$ nm (the length of pumping area), CoFeB thickness $h = 1$ nm (for perpendicular anisotropy), $\beta = 100$ fJ/Vm, for MgO $\varepsilon = 9.6$ and $\tan\delta = 10^{-5}$, tunnel resistance per area product $10^6 \Omega\mu m^2$ for $d = 2.2$ nm and increases in order per each 0.4 nm [6]. Then pumping electric field is $E \sim$ 1-1.5 V/nm, optimal MgO thickness $d \sim 2.6$ nm, energy consumption $\sim$ 3 aJ.